\documentclass[12pt]{article}
\usepackage{amsfonts,graphicx}
\newcommand{\beq}{\begin{equation}}
\newcommand{\eeq}{\end{equation}}
\newcommand{\beqa}{\begin{eqnarray}}
\newcommand{\eeqa}{\end{eqnarray}}
\newcommand{\beqar}{\begin{eqnarray*}}
\newcommand{\eeqar}{\end{eqnarray*}}

\newcommand{\norm}[1]{\raise.3ex\hbox{:}#1\raise.3ex\hbox{:}}

\newcommand{\labell}[1]{\label{#1}}

\begin{document}
{\titlepage{
\font\cmss=cmss10 \font\cmsss=cmss10 at 7pt

\vskip .1in \hfill hep-th/0306124

\hfill

\vspace{20pt}

\begin{center}
{\bf \large Supergravity Duals to the Noncommutative ${\cal N}=4$ SYM theory \\
      in the Infinite Momentum Frame}
\vspace*{1cm}

Hongsu Kim\footnote{E-mail: hongsu@astro.snu.ac.kr, @hepth.hanyang.ac.kr}

{\it Astronomy Program, SEES, Seoul National University}\\
{\it Seoul 151-742, Korea}\\
\vspace{2cm}

\textbf{Abstract}\end{center}
In this work the construction of supergravity duals to the noncommutative ${\cal N}=4$ SYM theory 
in the infinite momentum frame but with constant momentum density is attempted. 
In the absence of the content of noncommutativity, it has been known for some time that the 
previous $AdS_{5}/CFT_{4}$ correspondence should be replaced by the  
$K_{5}/CFT_{4}$ (with $K_{(p+2)}$ denoting the generalized Kaigorodov spacetime) 
correspondence with the pp-wave propagating on the BPS brane worldvolume. 
Interestingly enough, putting together the two contents, i.e., the introduction of noncommutativity 
and at the same time that of the pp-wave along the brane worldvolume, leads to quite nontrivial 
consequences such as the emergence of ``time-space'' noncommutativity in addition to the 
``space-space'' noncommutativity in the manifold on which the dual gauge theory is defined.
Taking the gravity decoupling limit, it has been realized that for small $u$, the solutions
all reduce to $K_{5}\times S^5$ geometry confirming our expectation that the IR dynamics of the
dual gauge theory should be unaffected by the noncommutativity while as $u\to \infty$, the
solutions start to deviate significantly from $K_{5}\times S^5$ limit indicating that the
UV dynamics of the dual gauge theory would be heavily distorted by the effect of noncommutativity.

\vfill \setcounter{page}{0} \setcounter{footnote}{0}
\newpage

\section{Introduction}

The celebrated $AdS/CFT$ correspondence \cite{ads-cft} is a conjectured (and by now fully tested)
equivalence between the two seemingly very different theories ; type IIB string theory on 
$AdS_{5}\times S^{5}$ and ${\cal N}=4$ SYM in $D=4$ with gauge group $SU(N)$. Since the advent
of this original duality conjecture, quite a few attempts have been made to extend it
to a larger context of gauge/gravity duality. The motivation of the present work can be thought
of as being along this line as it attempts to extend the original correspondence to the case when 
the dual CFT is in an infinitely-boosted frame on a noncommutative manifold.
To be more concrete, we would like to construct the extremal $D3$-brane solution with a superimposed
gravitational pp-wave in the presence of a $NSNS$ $B$-field background that is
supposed to be the supergravity dual (in the $AdS/CFT$ sense) to the non-commutative ${\cal N}=4$
super Yang-Mills (NCSYM) in the infinite momentum frame. 
Thus at this point, it seems relevant to address the issue of $AdS/CFT$ correspondence 
in cases where there is a pp-wave (for rather rigorous definition of the pp-wave, see appendix B) 
propagating along a direction longitudinal to a classical $p$-brane worldvolume \cite{clp, bcr}. 
And in this discussion, one has to carefully distinguish between the two cases ; 
the BPS-case and the non-BPS case. \\

First in the non-BPS case, the effect of including the gravitational pp-wave is locally
equivalent to performing a Lorentz boost along the propagation direction of the wave. 
(This point will be demonstrated explicitly in the appendix A.)
Furthermore, if the direction along which the pp-wave propagates is uncompactified, then the
equivalence is indeed valid globally, while if the propagation direction is wrapped on a circle,
it is valid only locally. For this reason, $p$-branes with superimposed pp-waves propagating
on their worldvolumes are often referred to as ``boosted'' $p$-branes. One, however, should bear in
mind that the global structure may not be precisely describable by a Lorentz boost.
In the BPS case, on the other hand, the inclusion of the pp-wave and the performance of a 
Lorentz boost along the propagation direction are {\it not} even locally equivalent.
This is because in the BPS limit, the Lorentz boost that relates the two metrics, one with
the pp-wave and the other without, becomes singular, corresponding to the ``infinite boost''
with the velocity approaching the speed of light. Thus in the BPS limit, one has essentially
two distinct configurations corresponding to the cases with and without the pp-wave which
are not even locally equivalent. \\

Next, an interesting consequence of this observation is the new structure of the near-horizon
geometry of the boosted $p$-branes. It is well-known that in the absence of the pp-wave, the
typical near-horizon geometries corresponding to the decoupling limit is of type
$AdS_{(p+2)}\times S^{D-(p+2)}$ for extremal, BPS $p$-branes such as BPS $M2$, $M5$ and $D3$-branes.
On the other hand, in the presence of the pp-wave propagating on a BPS $p$-brane, one can find
that the $AdS$-metric of near-horizon geometry is replaced by a new type of metric, which in
4-dimensions was first constructed long ago by Kaigorodov \cite{kaig} (and turned out to be of type $N$
in the Petrov classification). In other words, generally the near-horizon geometry of a BPS
$p$-brane with a pp-wave propagating on its worldvolume turns out to be of the type
$K_{(p+2)}\times S^{D-(p+2)}$ with $K_{(p+2)}$ denoting the $(p+2)$-dimensional generalization
of the Kaigorodov metric. Like $AdS$ geometry, the Kaigorodov spacetime is a homogeneous 
Einstein manifold, but they differ significantly in both their local and global structures.
In particular, although the Kaigorodov spacetimes approach $AdS$ locally at infinity, their 
boundaries are related to those of $AdS$ by an infinity Lorentz boost. This implies that the
boundary of the generalized Kaigorodov metric is in an infinite momentum frame and one can
moreover show that in the gravity decoupling limit, in order to maintain the structure of the
Kaigorodov metric, the momentun per unit $p$-volume must be held fixed. As a consequence, one
may conclude that the previous $AdS_{(p+2)}/CFT_{(p+1)}$ correspondence \cite{ads-cft}, the 
supergravity-boundary field theory duality that we associated to the extremal BPS $p$-brane solutions
should now be replaced by the  $K_{(p+2)}/CFT_{(p+1)}$ (with an infinite boost and a constant
momentum density) correspondence in the presence of the pp-wave propagating on the BPS
$p$-brane solutions. That is, in the spirit of gauge/gravity correspondence, it appears to be
natural to conjecture that string theory in $K_{(p+2)}\times S^{D-(p+2)}$ is dual to some
CFT in an infinitely-boosted frame, i.e., in the infinite momentum frame. Indeed, this conjecture
has been tested and actually confirmed to be true in a recent literature \cite{bcr}.
Next, in the case of non-BPS $p$-branes, the situation is somewhat different.
Namely, in the presence of the pp-wave propagating on non-BPS $p$-branes, the near-horizon 
geometries turn out to be of the type corresponding to the
[{\it Carter-Novotn$\acute{y}$-Horsk$\acute{y}$} spacetime]$\otimes S^{D-(p+2)}$. An interesting point,
however, in this non-BPS case is that there is locally no distinction between the case where
there is a superimposed pp-wave, and the case with no pp-wave. In fact, this can be attributed to 
the fact that a coordinate transformation (which is, as mentioned earlier, a Lorentz boost along
the wave propagation direction) allows the harmonic function associated with the
pp-wave to be set to unity. As a consequence, the local form of the 
Carter-Novotn$\acute{y}$-Horsk$\acute{y}$ (CNH) \cite{clp}
metrics remains the same whether or not a pp-wave is included in the original $p$-brane solution.
And the coordinate transformation (namely, the Lorentz boost) becomes singular in the extremal 
BPS-limit, which explains why there are two distinct cases in the BPS-situation leading either to 
the $AdS$ or else to the generalized Kaigorodov metrics after spherical reduction. To summarize,
unlike in the extremal case, in the non-extremal case, the pp-wave content can be erased by a
coordinate transformation (i.e., via a finite Lorentz boost). And for this reason, we shall in
the present work, focus exclusively on the extremal limit in which case there are two clearly
distinct correspondences : $AdS/CFT$ and $K/CFT$. \\

With this observations in mind, we now turn to the explicit construction of the ``extremal'' 
$(D1-D3)$ system
with a superimposed gravitational pp-wave. From
the rules for intersecting branes, it is known \cite{bergshoeff} that $M$-branes are parallely 
intersecting with the gravitational wave, i.e., $M2||W$ and $M5||W$. Thus 
starting with one of these configurations and applying the well-known duality web, one can deduce
a set of rules for putting $W$ (the pp-wave) on various p-branes in $D=10$ type II supergravity 
(SUGRA).
Therefore, following this standard procedure, we shall consider the following duality chain to 
obtain the $(D1-D3)$ system in the $NSNS$ $B$-field background with a pp-wave. 
\beq
M2||W \buildrel {\rm (KK)} \over \longrightarrow D2||W  
\buildrel {\rm (BMM-T)} \over \longrightarrow (D1-D3)||W
\eeq
where $KK$ denotes the Kaluza-Klein dimensional reduction from $D=11$ to $D=10$ along an
isometry direction and BMM-$T$ indicates the procedure suggested by Breckenridge, Michaud and
Myers \cite{bmm} to generally construct a $D_{p}-D_{(p+2)}$ bound state via a $T$-duality
transformation. Note that the endpoint of this duality chain is an ``electric'' solution
charged under electric $RR$ tensor field. If instead one follows the other duality chain, say,
\beq
M5||W  \buildrel {\rm (KK)} \over \longrightarrow D4||W 
\buildrel (T) \over \longrightarrow D3||W 
\buildrel (T) \over \longrightarrow D2||W 
\buildrel {\rm (BMM-T)} \over \longrightarrow  (D1-D3)||W,
\eeq
then one would end up with a ``magnetic'' $(D1-D3)$ system in the $NSNS$ $B$-field background 
with a pp-wave. And this is because the starting point is the $M5$-brane (with superimposed
pp-wave) which is the magnetic dual of the electric $M2$-brane solution of $D=11$ SUGRA. \\

As we mentioned, our primary concern in this work is the construction of supergravity duals to the 
noncommutative ${\cal N}=4$ SYM theory in the infinite momentum frame (but with constant momentum
density)\footnote{We were informed that some of the issues related to the ones addressed in the
present work also has been discussed in \cite{rcai}.}. 
Indeed, the construction of supergravity duals to the commutative ${\cal N}=4$ SYM theory 
in the infinite momentum frame has been discussed by Cvetic, Lu and Pope \cite{clp} and the 
supergravity duals to the noncommutative ${\cal N}=4$ SYM theory has been constructed by
Hashimoto and Itzhaki \cite{hi} and by Maldacena and Russo \cite{mr} some time ago.
Thus it may seem that the present work is a natural extension of these earlier works containing 
rather straightforwardly enlarged results. Interestingly enough, however, we shall see in a moment 
that putting the two contents, i.e., the introduction of noncommutativity and at the same time 
that of the pp-wave along the brane worldvolume, together leads to quite nontrivial consequences 
such as the emergence of ``time-space'' noncommutativity in addition to the ``space-space''
noncommutativity in the manifold on which the dual gauge theory is defined. \\

Lastly, perhaps it would be appropriate to distinguish the motivation and nature of the present
work from those of the recent development in the string theory on maximally supersymmetric 
pp-wave in association with the $AdS/CFT$ correspondence. Penrose \cite{penrose} has long ago pointed
out that, in the neighborhood of a null geodesic, all spacetimes locally have a plane wave as 
a limit. Indeed, plane waves are known examples of exact classical string vacuum. Thus by taking
the Penrose limit, any exact classical string vacuum can be related to the plane waves.
Then the great recent interest in the string/M-theory in the plane wave background resulted from
the realization that the maximally supersymmetric plane wave solution of type IIB SUGRA can thus
be obtained from the Penrose limit of the string vacuum solution, $AdS_{5}\times S^5$ \cite{pp1} 
and that
the superstring theory (particularly in the Green-Schwarz formalism with the choice of light-cone 
gauge) is exactly soluble in this plane wave background \cite{pp2}. 
Thus the motivation and the nature of this programme is to extend the $AdS/CFT$ correspondence to 
the regime of massive string states \cite{bmn} whereas those of the present work is to extend the
original $AdS/CFT$ correspondence still for the massless string spectrum to $K/CFT$ correspondence
with the boundary gauge theory being defined in the noncommutative manifold and moreover in the 
infinite momentum frame with a constant momentum density. We hope that the distinction between
the two is now clear.

\section{Near-horizon geometries of extremal/non-extremal M2-brane with a superimposed pp-wave}

We begin with the non-extremal $M2$-brane solution with a superimposed gravitational pp-wave
(see appendix A for a derivation)
\beqa
ds^2_{11} &=& H^{-2/3}\left[-K^{-1}fdt^2 + K \{dx_{1}+\coth \mu_{2}(K^{-1}-1)dt\}^2 
              + dx^2_{2}\right] + H^{1/3}\left[f^{-1}dr^2 + r^2 d\Omega^2_{7}\right], \nonumber \\
A_{[3]} &=& \coth \mu_{1}(H^{-1}-1) (dt\wedge dx_{1}\wedge dx_{2}) ~~~{\rm with} \\
H(r) &=& 1+{Q_{1}\over r^6}, ~~~K(r) = 1+{Q_{2}\over r^6},  ~~~f(r) = 1-{\mu\over r^6}, \nonumber \\
Q_{1} &=& \mu \sinh^2\mu_{1}, ~~~Q_{2} = \mu \sinh^2\mu_{2}, ~~~\mu = k\kappa^{4/3}_{11}  \nonumber
\eeqa
where $Q_{1}$ is the usual (electric) $RR$ charge and $Q_{2}$ is a new parameter representing
the momentum along the $x_{1}$-direction, i.e., $\mu_{2}$ parameterizes the Lorentz boost
factor as $\gamma = (1-\beta^2)^{-1/2} = \cosh\mu_{2}$ with $\beta = \tanh\mu_{2}$. And 
$\kappa_{11}$ denotes the 11-dimensional gravitational constant.
Then the associated extremal solution amounts to the limiting case 
\beq
\mu\to 0, ~~~\mu_{1}, ~\mu_{2}\to \infty ~~~{\rm with} ~~~Q_{1} = \mu \sinh^2\mu_{1},
~~~Q_{2} = \mu \sinh^2\mu_{2} ~~~{\rm kept ~~fixed}
\eeq
when the solution above becomes
\beqa
ds^2_{11} &=& H^{-2/3}\left[-K^{-1}dt^2 + K \{dx_{1}+(K^{-1}-1)dt\}^2 + dx^2_{2}\right]
              + H^{1/3}\left[dr^2 + r^2 d\Omega^2_{7}\right], \nonumber \\
A_{[3]} &=& (H^{-1}-1) (dt\wedge dx_{1}\wedge dx_{2}). 
\eeqa
In this section, we would like to explore the nature of the near-horizon geometries of both
extremal and non-extremal $M2$-brane solutions with the superimposed pp-wave in some
detail following \cite{clp}.

\subsection{Extremal solution}

We begin with the extremal case. Note first that under the coordinate transformation
\beq
t \to {3\over 2}t-{1\over 2}x_{1}, ~~~x_{1}\to {1\over 2}t + {1\over 2}x_{1}, \nonumber
\eeq
it follows $K\to (K-1) = Q_{2}/r^6$. This implies that the constant term ``$1$'' may be dropped
from $K(r)$ when it is more convenient to do so as it can always be removed via the coordinate
transformation given in eq.(6). Next, the {\it near-horizon}
region is defined to be $r\to 0$ where $H(r) \sim Q_{1}/r^6$. Thus the metric of the 
near-horizon geometry of the extremal $M2$-brane with superimposed pp-wave becomes
\beq
ds^2_{11} = Q_{1}^{-2/3}r^{4}\left[-K^{-1}dt^2 + K \{dx_{1}+(K^{-1}-1)dt\}^2 + dx^2_{2}\right]
              + Q_{1}^{1/3}{dr^2\over r^2} + Q_{1}^{1/3}d\Omega^2_{7}.
\eeq
Thus spacetime represented by this metric is a product $M_{4}\times S^{7}$ and particularly,
since the coefficient of the $S^7$ metric $d\Omega^2_{7}$ is a constant, $M_{4}$ here must
be an Einstein manifold with its metric being a solution of $D=4$ gravity with a pure
cosmological constant term
\beq
S_{4} = \int d^4x \sqrt{g_{4}}[R_{4} - 2\Lambda]
\eeq
where $\Lambda = -12Q_{1}^{-1/3}$. Now, upon the $S^7$ reduction and writing $r=e^{\rho}$,
the metric of the near-horizon geometry given above takes the form
\beq
ds_{4}^2 = Q_{1}^{1/3}\left[-e^{10\rho}dt^2 + e^{-2\rho}(dx_{1}+e^{6\rho}dt)^2
           +e^{4\rho}dx_{2}^2 + d\rho^2\right]
\eeq
where the charge parameters have been partly absorbed by rescaling the worldvolume
coordinates. It is straightforward to see that this is an homogeneous Einstein metric and
indeed it can be identified with the metric discovered first by Kaigorodov. We shall henceforth
denote it by $K_{4}$ and its generalization to arbitrary dimensions by $K_{n}$. Note also that
the $K_{4}$-metric has a 5-dim. isometry group (i.e., it possesses 5-Killing vectors) and it
preserves $1/4$ of the supersymmetry owned by the Minkowsi metric. Lastly, it might be of some
interest to compare this Kaigorodov metric with that of $AdS$ generally in $D=(n+3)$
dimensions. Consider the following family of metrics
\beq
ds_{D}^2 = -e^{2a\rho}dt^2 + e^{2b\rho}[dx + e^{(a-b)\rho}dt]^2
           +e^{2c\rho}dy^{i}dy^{i} + d\rho^2
\eeq
with $1\leq i\leq n$. Then there are actually two inequivalent solutions corresponding to the
metric having this ansatz that solves the Einstein equation $R_{\mu\nu}=\Lambda g_{\mu\nu}$
for a Einstein manifold and they are
\beqa
&&[AdS_{n+3}]  ~~~a = b = c = 2L, \nonumber \\
&&[K_{n+3}] ~~~a = (n+4)L, ~~b = -nL, ~~c = 2L \nonumber \\
&&{\rm with} ~~~L \equiv {1\over 2}\sqrt{{-\Lambda\over (n+2)}}. ~~~(\Lambda < 0) \nonumber
\eeqa

\subsection{Non-extremal solution}

Next, we turn to the study of the near-horizon geometry of the {\it non-extremal} $M2$-brane
with a superimposed pp-wave. In the near-horizon region, where 
$H(r) \sim Q_{1}/r^6 = (k\kappa^{4/3}_{11})\sinh^2\mu_{1}/r^6$, the metric of non-extremal
$M2$-brane with a superimposed pp-wave becomes
\beqa
ds^2_{11} &=& Q_{1}^{-2/3}r^{4}\left[-K^{-1}fdt^2 + K \{dx_{1}+\coth\mu_{2}(K^{-1}-1)dt\}^2 
              + dx^2_{2}\right] \\
           &+& Q_{1}^{1/3}f^{-1}{dr^2\over r^2} + Q_{1}^{1/3}d\Omega^2_{7} \nonumber
\eeqa
which is again of the metric form for $M_{4}\times S^7$ with $M_{4}$ metric being a solution
of $D=4$ pure gravity with only a cosmological constant term represented by the action
\beq
S_{4} = \int d^4x \sqrt{g_{4}}[R_{4} - 2\Lambda]
\eeq
where $\Lambda = -12Q^{-1/3}_{1} = -12(k\kappa^{4/3}_{11} \sinh^2\mu_{1})^{-1/3}$.
As before, we take $S^7$ as the ``internal'' sphere having the metric
$ds^2_{7} = Q^{1/3}_{1}d\Omega^2_{7} = (k\kappa^{4/3}_{11} \sinh^2\mu_{1})^{1/3}
d\Omega^2_{7}$ or equivalently having the radius 
$R_{7} = Q^{1/6}_{1} = (k\kappa^{4/3}_{11} \sinh^2\mu_{1})^{1/6}$. Then upon the
$S^7$-reduction and writing $r = e^{\rho}$, the metric of the near-horizon geometry given above
takes the form
\beq
ds^2_{4} = Q_{1}^{-2/3}e^{4\rho}\left[-K^{-1}fdt^2 + K \{dx_{1}+\coth \mu_{2}(K^{-1}-1)dt\}^2 
              + dx^2_{2}\right] + Q_{1}^{1/3}f^{-1}d\rho^2 
\labell{ssusy}
\eeq
with $K(\rho) = 1+k\sinh^2\mu_{2}e^{-6\rho}$, $f(\rho) = 1-ke^{-6\rho}$.
(Here, we have set the gravitational constant $\kappa_{11}$ to unity for convenience.)
This is indeed a metric for an Einstein manifold found by Carter and by Novotn$\acute{y}$ and
Horsk$\acute{y}$ (CNH).  In the asymptotic region where $r\to \infty$, we have $f(r)\to 1$ and
hence this CNH metric goes over to the Kaigorodov metric discussed earlier. 
Lastly, we comment on the generalization of this (originally 4-dimensional) CNH metric to
arbitrary dimensions for later use. First, the generalized CNH metric that typically arises
in the spherical $S^{D-(p+2)}$-reduction of the non-extremal $p$-brane with a superimposed
pp-wave is given by
\beq
ds^2_{p+2} = c_{1}e^{2{\tilde{d}\over d}\rho}\left[-K^{-1}fdt^2 + K \{dx_{1}
               +\coth \mu_{2}(K^{-1}-1)dt\}^2 + dy^{i}dy^{i}\right] + c_{2}f^{-1}d\rho^2 
\label{phisol}
\eeq
with $K(\rho) = 1+k\sinh^2 \mu_{2}e^{-\tilde{d}\rho}$, $f(\rho) = 1-ke^{-\tilde{d}\rho}$ and
$d = (p+1)$, $(\tilde{d}+2) = D-d$. Or more generally, the solution to the Einstein equation
$R_{\mu\nu} = \Lambda g_{\mu\nu}$ in $D = (n+3)$ that represents the generalization of the 
above CNH metric is given by
\beq
ds_{D}^2 = -e^{2a\rho}fdt^2 + e^{2b\rho}[dx + e^{(a-b)\rho}dt]^2
           +e^{2c\rho}dy^{i}dy^{i} + f^{-1}d\rho^2
\eeq
where $1\leq i\leq n$, $f(\rho) = 1-ke^{-(a-b)\rho}$ and
\beq
a = (n+4)L, ~~b = -nL, ~~c = 2L,
~~~{\rm with} ~~L \equiv {1\over 2}\sqrt{{-\Lambda\over (n+2)}} ~~(\Lambda < 0). \nonumber
\labell{action}
\eeq
Note also that this generalized CNH metric can also be put in the form 
\beqa
ds_{D}^2 = e^{-2nL\rho}dx^2 + e^{4L\rho}(2dxdt + kdt^2 + dy^{i}dy^{i}) +
           \left[1 - ke^{-2(n+2)L\rho}\right]^{-1}d\rho^2.
\labell{d1d3enbound}
\eeqa
Both of these expressions for the generalized CNH metric given in eqs.(15) and (17) reduce to that 
the generalized Kaigorodov metric given in eq.(10) for $k = 0$. This completes the study of near-horizon 
geometries of $M2$-brane with a superimposed pp-wave. In the following section, we get back
to our main task of constructing the extremal $(D1-D3)$ system with a superimposed pp-wave in the 
presence of the $NSNS$ $B$-field.  

\section{Construction of $(D1-D3)$ system with a superimposed pp-wave}

\subsection{Construction of the solution}

We now describe our strategy briefly. From the rules for intersecting branes, it is known \cite{bergshoeff} 
that $M$-branes are parallely intersecting with the $M$-wave, i.e., $(1|M2,W)$ (or $M2||W$ for short)
and $(1|M5,W)$ (or $M5||W$ for short).
Thus (I) we shall start with the extremal $M2$-brane solution with a
superimposed gravitational pp-wave, $M2||W$ and then perform a Kaluza-Klein (KK) dimensional reduction 
to $D=10$ along a U(1)-isometry direction which is chosen to be a coordinate transverse to the
$M2$-brane to get $D2||W$ in IIA theory. Then next, (II) we shall proceed with the procedure suggested
by Breckenridge, Michaud and Myers (BMM) \cite{bmm} to finally obtain $(D1-D3)||W$ via a $T$-dual 
transformation :  \\

(I) $M2||W \buildrel {\rm (KK)} \over \longrightarrow  D2||W$ \\

Consider the extremal $M2$-brane solution with a superimposed pp-wave and its KK reduction,
\beqa
ds^2_{11} &=& H^{-2/3}\left[-K^{-1}dt^2 + K \{dx_{1}+(K^{-1}-1)dt\}^2 + dx^2_{2}\right]
              + H^{1/3}\left[\sum^{10}_{m=3}dx^2_{m}\right] \nonumber \\
&=& e^{-{2\over 3}\phi}ds^2_{10} + e^{{4\over 3}\phi}(dy + A^{(1)}_{\mu}dx^{\mu})^2, \nonumber \\
A_{[3]} &=& (H^{-1}-1) dt\wedge dx_{1}\wedge dx_{2} \\
&=& {1\over 3!}A^{(3)}_{\mu\nu\lambda}dx^{\mu}\wedge dx^{\nu}\wedge dx^{\lambda} + 
    {1\over 2!}B^{(2)}_{\mu\nu}dx^{\mu}\wedge dx^{\nu}\wedge dy \nonumber
\eeqa
where $A^{(3)}$ and $B^{(2)}$ denote respectively a 3-form $RR$ and a 2-form $NSNS$ potential and
$A^{(1)}$ is the KK gauge field. $H(r)$ and $K(r)$ are as given before.
Reduction to $D=10$ along a direction transverse to the $M2$-brane amounts to choosing, say,
$y=x_{3}$. This then implies that we should identify $e^{{4/3}\phi} = H^{1/3}$ which, in turn, yields
\beq
e^{\phi} = H^{1/4}, ~~A^{(1)}_{\mu} = 0, ~~B^{(2)}_{\mu\nu} = 0,
~~A^{(3)}_{\mu\nu\lambda} = \{A^{(3)}_{tx_{1}x_{2}}=(H^{-1} - 1)\}.
\eeq
Thus the result is a $D2$-brane solution in $D=10$ type IIA SUGRA with a superimposed pp-wave given by
\beqa
ds^2_{10} &=& H^{-1/2}\left[-K^{-1}dt^2 + K \{dx_{1}+(K^{-1}-1)dt\}^2 + dx^2_{2}\right]
              + H^{1/2}\left[\sum^{9}_{m=3}dx^2_{m}\right], \nonumber \\
A_{[3]} &=& {1\over g_{s}}(H^{-1}-1) dt\wedge dx_{1}\wedge dx_{2}, \\
e^{2\phi} &=& g^2_{s}H^{1/2} \nonumber
\eeqa
where $g_{s}$ denotes the string coupling representing $g_{s}=e^{\phi_{\infty}}$.	\\

(II) $D2||W \buildrel {\rm (BMM-T)} \over \longrightarrow  (D1-D3)||W$  \\

Finally, following the procedure suggested by Breckenridge, Michaud and Myers (BMM)\cite{bmm}, 
we now construct the bound state of extremal $D1-D3$ system with a superimposed pp-wave.
For the case at hand, the suggested procedure of BMM to construct a $D1-D3$ system in the
$NSNS$ $B$-field background consists of the following 3-steps. \\

(i) Delocalize (or smear out) the $D2$-brane (oriented in $(x_{1}-x_{2})$ plane) in one of the
transverse (say, $x_{3}$) directions. \\

(ii) Perform a rotation on the delocalized $D2$-brane in $(x_{2}-x_{3})$ plane namely, we make it
to be `tilted' in  $(x_{2}-x_{3})$ plane. \\

(iii) Apply $T$-duality on (rotated) $\tilde{x}_{3}$-direction. \\

In other words, we begin with the extremal $D2$-brane plus wave given above but oriented instead
at an angle in $(x_{2}-x_{3})$ plane and then apply $T$-duality on $x_{3}$ to find a solution
describing the bound state of an extremal $D1$ and $D3$-brane with a pp-wave.
Thus we start by rewriting the $D2||W$ solution given above as
\beqa
ds^2_{10} &=& H^{1/2}\left[{-K^{-1}dt^2 + K \{dx_{1}+(K^{-1}-1)dt\}^2 + dx^2_{2}\over
              H} + \sum^{9}_{m=3}dx^2_{m}\right], \nonumber \\
A_{[3]} &=& {1\over g_{s}}(H^{-1}-1) dt\wedge dx_{1}\wedge dx_{2}, \\
e^{2\phi} &=& g^2_{s}H^{1/2} = g^2_{s}\left[1 + {Q_{1}\over r^5}\right]^{1/2}. \nonumber
\eeqa
Recall that $H(r)$ here is a harmonic function in the transverse coordinates which solves the
Poisson's equation with some delta function source. According to the suggested prescription of BMM,
one needs a slightly different harmonic function $H(r)$ (and $K(r)$ as well for the case at hand
where we consider the superposition of a pp-wave) in order to `delocalize' the present extremal
$D2$-brane (with a pp-wave) in one of the transverse directions, say, $x_{3}$. And then they pointed
out that this can be done in at least two different ways. Firstly, the delta function source can be
chosen so that $\partial_{i}\partial^{i}H = -5Q_{1}A_{6}\prod^{9}_{i=3}\delta(x_{i})$
(where $A_{n}$ denotes the area of a unit $n$-sphere) and the delocalization of the $D2$-brane
can be achieved by following the so-called `vertical reduction' approach. Namely, one adds an
infinite number of identical sources in a periodic array along the $x_{3}$-axis. Then a smeared
solution may be extracted from the long-range fields, for which the $x_{3}$-dependence is
exponentially suppressed. The second approach, which may be termed, `vertical oxidation', consists in
simply replacing the above 7-dimensional delta function source by that of a line source extending
along $x_{3}$, i.e., $\partial_{i}\partial^{i}H = -4Q_{1}A_{5}\prod^{9}_{i=4}\delta(x_{i})$.
Whichever method one may employ, the number of dimensions transverse to the ``smeared-out'' 
$D2$-brane becomes `effectively' only 6 rather than 7, i.e.,
\beqa
ds^2_{10} &=& H^{1/2}\left[{-K^{-1}dt^2 + K \{dx_{1}+(K^{-1}-1)dt\}^2 + dx^2_{2}\over
              H} + dx^2_{3} + \sum^{9}_{m=4}dx^2_{m}\right], \nonumber \\
H(r) &=& 1 + {Q_{1}\over r^4},  ~~K(r) = 1 + {Q_{2}\over r^4} ~~{\rm and}
~~r^2 = \sum^{9}_{i=4}x^2_{i}. 
\eeqa
Then the form of the antisymmetric $RR$ tensor potential
$A_{[3]} = {1\over g_{s}}(H^{-1}-1) dt\wedge dx_{1}\wedge dx_{2}$ reveals that we now have 
a $D2$-brane oriented
along $(x_{1}-x_{2})$ plane and smeared out in $x_{3}$-direction. 
We now consider performing a rotation
on our delocalized $D2$-brane. Note, however, that since our $D2$-brane was originally extended
in  $(x_{1}-x_{2})$ plane, we may have two {\it inequivalent} options : \\

(A) the rotation in $(x_{2}-x_{3})$ plane with $x_{2}$ being the `spectator' direction (with respect
to the superimposed pp-wave) or \\

(B) the rotation in $(x_{1}-x_{3})$ plane now with $x_{1}$ being the `boost' (or wave propagation)
direction. \\

Obviously, the option (A) would not distort the pp-wave propagating on the brane worldvolume and we
discuss this case first and then the option (B) later on. \\
Namely the rotation
\beqa
\left\{ \begin{array}{clcr}
  dx_{3} = \cos \varphi d\tilde{x}_{3} - \sin \varphi d\tilde{x}_{2}, & \;\;\;  \\
  dx_{2} = \sin \varphi d\tilde{x}_{3} + \cos \varphi d\tilde{x}_{2} & \;\;\; 
\end{array} \right.
\label{nomono}
\eeqa
with $\varphi $ being the angle between $\tilde{x}_{2}$-axis and $x_{2}$-axis,
takes the $D2$-brane plus the pp-wave solution given above to
\beqa
ds^2_{10} &=& H^{1/2}\left[{-K^{-1}dt^2 + K \{dx_{1}+(K^{-1}-1)dt\}^2 \over
H} + \left({\cos^2 \varphi \over H} + \sin^2 \varphi \right)d\tilde{x}^{2}_{2} \right. \nonumber \\
              &&\left. + \left({\sin^2 \varphi \over H} + \cos^2 \varphi \right)d\tilde{x}^{2}_{3}  +  
              2\cos \varphi \sin \varphi \left({1\over H} - 1\right)d\tilde{x}_{2} 
              d\tilde{x}_{3} + \sum^{9}_{m=4}dx^2_{m}\right], \nonumber \\
A_{[3]} &=& {1\over g_{s}}(H^{-1}-1) dt\wedge dx_{1}\wedge 
            (\cos \varphi d\tilde{x}_{2} + \sin \varphi d\tilde{x}_{3}), \\
e^{2\phi} &=& g^2_{s}H^{1/2} = g^2_{s}\left[1 + {Q_{1}\over r^4}\right]^{1/2}. \nonumber
\labell{Nnrel}
\eeqa
Lastly, applying the generalized Buscher's $T$-duality\footnote{The explicit generalized Bucher's
$T$-duality rules employed in this work can be found for instance in \cite{tduals}.} 
on $\tilde{x}_{3}$, we end up with
\beqa
ds^2_{10} &=& H^{1/2}\left[{-K^{-1}dt^2 + K \{dx_{1}+(K^{-1}-1)dt\}^2 \over
              H} + {d\tilde{x}^{2}_{2} + d\tilde{x}^{2}_{3}\over 1+(H-1)\cos^2 \varphi} 
              + \sum^{9}_{m=4}dx^2_{m}\right], \nonumber \\
B_{[2]} &=& {1\over g_{s}}{(H-1)\cos \varphi \sin \varphi \over 1+(H-1)\cos^2 \varphi}
            (d\tilde{x}_{2}\wedge d\tilde{x}_{3}), \nonumber \\
A_{[2]} &=& {1\over g_{s}}(H^{-1}-1) \sin \varphi (dt\wedge dx_{1}), \\
A_{[4]} &=& {1\over g_{s}}{(H-1)\cos \varphi \over 1+(H-1)\cos^2 \varphi}
            (dt\wedge dx_{1}\wedge d\tilde{x}_{2}\wedge d\tilde{x}_{3}), \nonumber \\
e^{2\phi} &=& g^2_{s}{H\over 1+(H-1)\cos^2 \varphi} \nonumber
\labell{physR}
\eeqa
with $H(r)$, $K(r)$ as given before. Note that it is evident from the emergence of $RR$
potentials $A_{[2]}$ and $A_{[4]}$ that we indeed have a bound state of a $D1$ and
$D3$-branes. Moreover, this solution manifests itself as representing a $D1$-brane ``dissolved''
in the $D3$-brane and having a superimposed pp-wave propagating on its worldvolume. \\
As has been noted in the introduction, one can obtain 
a ``magnetic'' $(D1-D3)$ system in the $NSNS$ $B$-field background 
with a pp-wave as well by starting with the $M5$-brane (with superimposed
pp-wave) and following the duality chain given in eq.(2). Then the magnetically $RR$-charged
solution turns out to be the same as the electrically-charged solution given above except that
now the $RR$ tensor field strengths are given, instead of $F^{e}_{[3]}=dA_{[2]}$ and
$F^{e}_{[5]}=dA_{[4]}$, by
\beqa
&&F^{m}_{[7]} = ^{*}F^{e}_{[3]} = ^{*}(dA_{[2]}) = -{4Q_{1}\sin \varphi \over g_{s}r^{6}
[1+(H-1)\cos^2 \varphi]} \times \\
&&[x_{4}(dx_{2}\wedge dx_{3}\wedge dx_{5}\wedge dx_{6}\wedge dx_{7}\wedge 
dx_{8}\wedge dx_{9}) - x_{5}(dx_{2}\wedge dx_{3}\wedge dx_{4}\wedge dx_{6}\wedge dx_{7}\wedge 
dx_{8}\wedge dx_{9}) \nonumber \\
&&+ x_{6}(dx_{2}\wedge dx_{3}\wedge dx_{4}\wedge dx_{5}\wedge dx_{7}\wedge 
dx_{8}\wedge dx_{9}) - x_{7}(dx_{2}\wedge dx_{3}\wedge dx_{4}\wedge dx_{5}\wedge dx_{6}\wedge 
dx_{8}\wedge dx_{9}) \nonumber \\
&&+ x_{8}(dx_{2}\wedge dx_{3}\wedge dx_{4}\wedge dx_{5}\wedge dx_{6}\wedge 
dx_{7}\wedge dx_{9}) - x_{9}(dx_{2}\wedge dx_{3}\wedge dx_{4}\wedge dx_{5}\wedge dx_{6}\wedge 
dx_{7}\wedge dx_{8})], \nonumber \\
&&F^{m}_{[5]} = ^{*}F^{e}_{[5]} = ^{*}(dA_{[4]}) = {4Q_{1}H\cos \varphi \over g_{s}r^{6}
[1+(H-1)\cos^2 \varphi]} \times \\
&&[x_{4}(dx_{5}\wedge dx_{6}\wedge dx_{7}\wedge dx_{8}\wedge dx_{9})
- x_{5}(dx_{4}\wedge dx_{6}\wedge dx_{7}\wedge dx_{8}\wedge dx_{9})  \nonumber \\
&&+ x_{6}(dx_{4}\wedge dx_{5}\wedge dx_{7}\wedge dx_{8}\wedge dx_{9})
- x_{7}(dx_{4}\wedge dx_{5}\wedge dx_{6}\wedge dx_{8}\wedge dx_{9}) \nonumber \\
&&+ x_{8}(dx_{4}\wedge dx_{5}\wedge dx_{6}\wedge dx_{7}\wedge dx_{9})
- x_{9}(dx_{4}\wedge dx_{5}\wedge dx_{6}\wedge dx_{7}\wedge dx_{8})] \nonumber 
\eeqa
where the Hodge dual is taken with respect to the metric solution given in eq.(25) above.
Note that $F^{m}_{[5]} \neq F^{e}_{[5]}$ and hence $F^{e}_{[5]} \neq ^{*}F^{e}_{[5]}$, namely
the 5-form $RR$ field strength is not self-dual. \\

Next, we turn to the option (B) in which the $D2$-brane (delocalized in the $x_{3}$-direction)
is to be rotated in $(x_{1}-x_{3})$ plane with $x_{1}$ being the boost (i.e., wave
propagation) direction. Namely, upon the rotation
\beqa
\left\{ \begin{array}{clcr}
  dx_{3} = \cos \varphi d\tilde{x}_{3} - \sin \varphi d\tilde{x}_{1}, & \;\;\;  \\
  dx_{1} = \sin \varphi d\tilde{x}_{3} + \cos \varphi d\tilde{x}_{1} & \;\;\; 
\end{array} \right.
\label{di}
\eeqa
with this time $\varphi $ being the angle between $\tilde{x}_{1}$-axis and $x_{1}$-axis,
the $D2$-brane plus the pp-wave solution given above becomes
\beqa
ds^2_{10} &=& H^{1/2}\left[{1\over H}\{-(2-K)dt^2 + 2(1-K)dt(\cos \varphi d\tilde{x}_{1}
              + \sin \varphi d\tilde{x}_{3})\} + {dx^2_{2}\over H} \right. \nonumber \\
              &&\left. + \left({K\over H}\cos^2 \varphi + \sin^2 \varphi \right)d\tilde{x}^{2}_{1} +  
              \left({K\over H}\sin^2 \varphi + \cos^2 \varphi \right)d\tilde{x}^{2}_{3} \right. \nonumber \\ 
              &&\left. +  2\cos \varphi \sin \varphi \left({K\over H} - 1\right)d\tilde{x}_{1} 
              d\tilde{x}_{3} + \sum^{9}_{m=4}dx^2_{m}\right], \nonumber \\
A_{[3]} &=& {1\over g_{s}}(H^{-1}-1) \{dt\wedge 
            (\cos \varphi d\tilde{x}_{1} + \sin \varphi d\tilde{x}_{3})\wedge dx_{2}\}, \\
e^{2\phi} &=& g^2_{s}H^{1/2} = g^2_{s}\left[1 + {Q_{1}\over r^4}\right]^{1/2}. \nonumber
\labell{tri}
\eeqa
Lastly, applying the generalized Buscher's $T$-duality on $\tilde{x}_{3}$, we are left with
\beqa
ds^2_{10} &=& H^{1/2}\left[-{K^{-1}dt^2\over H} + {K \{d\tilde{x}_{1}+\cos \varphi (K^{-1}-1)dt\}^2 
              \over K+(H-K)\cos^2 \varphi} + {dx^2_{2}\over H} \right. \nonumber \\
              &&\left. + {d\tilde{x}^{2}_{3}\over K+(H-K)\cos^2 \varphi} 
              + \sum^{9}_{m=4}dx^2_{m}\right], \nonumber \\
B_{[2]} &=& {1\over g_{s}}{(K-1)\sin \varphi \over K+(H-K)\cos^2 \varphi}
            (dt\wedge d\tilde{x}_{3}) +
            {1\over g_{s}}{(H-K)\cos \varphi \sin \varphi \over K+(H-K)\cos^2 \varphi}
            (d\tilde{x}_{1}\wedge d\tilde{x}_{3}), \nonumber \\
A_{[2]} &=& {1\over g_{s}}(H^{-1}-1) \sin \varphi (dt\wedge dx_{2}), \\
A_{[4]} &=& {1\over g_{s}}{(H-1)\cos \varphi \over K+(H-K)\cos^2 \varphi}
            (dt\wedge d\tilde{x}_{1}\wedge dx_{2}\wedge d\tilde{x}_{3}), \nonumber \\
e^{2\phi} &=& g^2_{s}{H\over K+(H-K)\cos^2 \varphi} \nonumber
\labell{quad}
\eeqa
with again $H(r)$, $K(r)$ as given before. \\
Next, the associated magnetically $RR$-charged solution again turns out to be the same as the 
electrically-charged solution given above except that
the $RR$ tensor field strengths are given, instead of $F^{e}_{[3]}=dA_{[2]}$ and
$F^{e}_{[5]}=dA_{[4]}$, by
\beqa
F^{m}_{[7]} &=& ^{*}F^{e}_{[3]} = ^{*}(dA_{[2]}) = -{4Q_{1}\sin \varphi \over g_{s}r^{6}
[K+(H-K)\cos^2 \varphi]} \times \\
&[&x_{4}(dx_{3}\wedge dx_{5}\wedge dx_{6}\wedge dx_{7}\wedge 
dx_{8}\wedge dx_{9}) - x_{5}(dx_{3}\wedge dx_{4}\wedge dx_{6}\wedge dx_{7}\wedge 
dx_{8}\wedge dx_{9}) \nonumber \\
&+& x_{6}(dx_{3}\wedge dx_{4}\wedge dx_{5}\wedge dx_{7}\wedge 
dx_{8}\wedge dx_{9}) - x_{7}(dx_{3}\wedge dx_{4}\wedge dx_{5}\wedge dx_{6}\wedge 
dx_{8}\wedge dx_{9}) \nonumber \\
&+& x_{8}(dx_{3}\wedge dx_{4}\wedge dx_{5}\wedge dx_{6}\wedge 
dx_{7}\wedge dx_{9}) - x_{9}(dx_{3}\wedge dx_{4}\wedge dx_{5}\wedge dx_{6}\wedge 
dx_{7}\wedge dx_{8})] \nonumber \\
&&\wedge [(1-K)\cos \varphi dt + Kdx_{1}], \nonumber \\
F^{m}_{[5]} &=& ^{*}F^{e}_{[5]} = ^{*}(dA_{[4]})   \\
&=& {H\cos \varphi \over [K+(H-K)\cos^2 \varphi]} 
\left\{[K+(1-K)\cos^2 \varphi]{4Q_{1}\over g_{s}r^6}
+(1-H)\sin^2 \varphi {4Q_{2}\over g_{s}r^6}\right\} \times \nonumber \\
&[&x_{4}(dx_{5}\wedge dx_{6}\wedge dx_{7}\wedge dx_{8}\wedge dx_{9})
- x_{5}(dx_{4}\wedge dx_{6}\wedge dx_{7}\wedge dx_{8}\wedge dx_{9})  \nonumber \\
&+& x_{6}(dx_{4}\wedge dx_{5}\wedge dx_{7}\wedge dx_{8}\wedge dx_{9})
- x_{7}(dx_{4}\wedge dx_{5}\wedge dx_{6}\wedge dx_{8}\wedge dx_{9}) \nonumber \\
&+& x_{8}(dx_{4}\wedge dx_{5}\wedge dx_{6}\wedge dx_{7}\wedge dx_{9})
- x_{9}(dx_{4}\wedge dx_{5}\wedge dx_{6}\wedge dx_{7}\wedge dx_{8})] \nonumber 
\eeqa
where the Hodge dual is taken with respect to the metric solution given in eq.(30).
Note again that $F^{m}_{[5]} \neq F^{e}_{[5]}$ and hence $F^{e}_{[5]} \neq ^{*}F^{e}_{[5]}$, namely
the 5-form $RR$ field strength is not self-dual. 

\subsection{Nature of the solution}

First of all, again it is evident from the 
emergence of $RR$ potentials $A_{[2]}$ and $A_{[4]}$ that we are left with a bound state of
a $D1$ and $D3$-branes. Moreover, this solution still appears to represent a $D1$-brane 
``dissolved'' in a $D3$-brane with a superimposed pp-wave propagating in $\tilde{x}_{1}$-direction
which is tilted from the original $x_{1}$-direction by an angle $\varphi$. Particularly, 
a remarkable feature of the solution corresponding to option (B) that can be contrasted from 
that of our previous solution corresponding
to option (A) is that now we have the non-vanishing $NSNS$ $B$-field component 
$B_{t\tilde{x}_{3}}$ as well as the usually expected component $B_{\tilde{x}_{1}\tilde{x}_{3}}$.
Note that we shall eventually propose that these solutions are the dual supergravity description
of noncommutative SYM at large coupling and in the {\it infinite-momentum-frame} particularly
if we consider the gravity decoupling limits of their extremal versions. And of course this
interpretation is based on the $K_{(p+2)}/CFT_{(p+1)}$ (in the infinite-momentum-frame but with
a constant momentum density) correspondence we discussed earlier in the introduction. In this
spirit, the emergence of non-vanishing components $B_{t\tilde{x}_{3}}$ and
$B_{\tilde{x}_{1}\tilde{x}_{3}}$ in the dual supergravity solution corresponding to option (B)
implies that its dual SYM theory at large coupling should be defined on a manifold consisting
of the two noncommutative hypersurfaces $(t - \tilde{x}_{3})$ and 
$(\tilde{x}_{1} - \tilde{x}_{3})$ planes. Namely, we now ended up with both ``time-space'' and
``space-space'' noncommutativity in option (B) in contrast to option (A) where one was left
with just ``space-space'' noncommutativity. \\
Although it may, at first sight, seem quite a surprise, it, on second thought, was rather an
expected result. That is to say, first notice that it is the ``tilting'' procedure of the
delocalized $D2$-brane that essentially generates the $NSNS$ $B$-field components upon performing
the $T$-duality. Then in option (A), the rotation is done in $(x_{2}-x_{3})$ plane with $x_{2}$
being a spectator direction with respect to the propagating pp-wave on the brane. Thus the 
$T$-duality can at most generates the component $B_{\tilde{x}_{2}\tilde{x}_{3}}$. In option (B),
on the other hand, the rotation is performed in $(\tilde{x}_{1} - \tilde{x}_{3})$ plane instead
with $x_{1}$ now being the wave propagation direction. As a result, due to the non-vanishing
metric component $g_{t\tilde{x}_{1}}$ this time, the $T$-duality turns out to generate 
non-zero component $B_{t\tilde{x}_{3}}$ as well as $B_{\tilde{x}_{1}\tilde{x}_{3}}$. And this is
why we have both time-space and space-space noncommutativity in its dual SYM theory for
option (B). Namely, it is the non-trivial role played by the superimposed pp-wave that leads to
the full noncommutativity in its dual SYM theory. Lastly, we also note that if $Q_{2}=0$
(and hence $K(r)=1$), namely in the absence of the superimposed pp-wave, both of these
$(D1-D3)$ bound state solutions given above correctly reduce to that of Hashimoto and Itzhaki \cite{hi}
or of Maldacena and Russo \cite{mr} with two (spectator) longitudinal directions $x_{1}$, $x_{2}$ which
can now be freely interchanged.

\section{Decoupling Limits}

The metric sector of the extremal $(D1-D3)$ bound state solutions with a superimposed pp-wave
given above all asymptote to the 10-dimensional flat spacetime as $r\rightarrow \infty $.
Very near the horizon at $r = 0$, on the other hand, they nearly look like $K_{5}\times S^{5}$
with $K_{5}$ being the 5-dimensional generalisation of the ``Kaigorodov'' metric we discussed
in some detail earlier. And the throat connecting these two asymptotic regions contains 
non-zero $NSNS$ and $RR$ fields. Thus on the boundary, we would have the ${\cal N}=4$ SYM theory
in the infinitely-boosted frame but with constant and finite momentum density in the spirit of
$K_{(p+2)}/CFT_{(p+1)}$ correspondence in the presence of the pp-wave propagating on the 
extremal $p$-brane worldvolume. Therefore, we now elaborate on this point. In order eventually
to have noncommutative SYM theory in the infinitely-boosted frame (living in the worldvolume
of $(D1-D3)$ system placed on the boundary near the horizon) on the boundary of $K_{5}$, we
take the following ordinary field theory decoupling limit : 
\beqa
\alpha' &\rightarrow& 0, ~~~\tan \varphi = {\tilde{b}\over \alpha'},  \\
r &=& \alpha' R^2 u, ~~~g_{s} = {\alpha'\over \tilde{b}}\hat{g}_{s} \nonumber
\labell{chargeeq}
\eeqa
where $R^{4} = 4\pi g_{s}N$ and
$x_{0,1} = \tilde{x}_{0,1}$, $x_{2,3} = {\alpha'\over \tilde{b}}\tilde{x}_{2,3}$ for the
solution corresponding to option (A) and 
$x_{0,2} = \tilde{x}_{0,2}$, $x_{1,3} = {\alpha'\over \tilde{b}}\tilde{x}_{1,3}$ for the
solution corresponding to option (B), with $u, ~\hat{g}_{s}, ~\tilde{x}_{\mu}, ~\tilde{b}$
being kept fixed. Here the factor of $\tilde{b}$ has been introduced for later convenience.
And $\hat{g}_{s}$ here is the value of string coupling in the IR regime. Now in this decoupling
limit, the extremal solution corresponding to option (A) becomes
\beqa
ds^2_{10} &=& \alpha' R^2\left[u^2 \{-K^{-1}d\tilde{t}^2 + K(d\tilde{x}_{1}+(K^{-1}-1)d\tilde{t})^2\} 
              + u^2\hat{h}(d\tilde{x}^{2}_{2} + d\tilde{x}^{2}_{3}) 
              + {du^2 \over u^2} + d\Omega^2_{5}\right], \nonumber \\
B_{[2]} &=& B_{\infty}{(\alpha'^2 R^4 u^4-1)\over (1+a^4 u^4)}
            (d\tilde{x}_{2}\wedge d\tilde{x}_{3}), \nonumber \\
A_{[2]} &=& -{1\over g_{s}}\left({\tilde{b}\over \alpha'}\right)(\alpha'^2 R^4 u^4-1)
            (d\tilde{t}\wedge d\tilde{x}_{1}), \\
A_{[4]} &=& {\hat{h}\over g_{s}}(\alpha'^2 R^4 u^4-1)
            (d\tilde{t}\wedge d\tilde{x}_{1}\wedge d\tilde{x}_{2}\wedge d\tilde{x}_{3}), 
             \nonumber \\
e^{2\phi} &=& \hat{g}^2_{s}\hat{h} \nonumber
\eeqa
where $K=Q_{2}/(\alpha' R^2 u)^4$, $\hat{h}=1/(1+a^4 u^4)$ and $B_{\infty}=\alpha'/\tilde{b}
= \alpha'(R^2/a^2)$ with $a=R\sqrt{\tilde{b}}$. \\
Next, the decoupling limit of the extremal solution corresponding to option (B) reads
\beqa
ds^2_{10} &=& \alpha' R^2\left[u^2(-K^{-1}d\tilde{t}^2 + d\tilde{x}^{2}_{2})
              + u^2\hat{h} \{K[d\tilde{x}_{1}+(K^{-1}-1)d\tilde{t}]^2 + d\tilde{x}^{2}_{3}\} 
              + {du^2 \over u^2} + d\Omega^2_{5}\right], \nonumber \\
B_{[2]} &=& B_{\infty}{(1-K)a^4 u^4\over (1+Ka^4 u^4)}(d\tilde{t}\wedge d\tilde{x}_{3})
            + B_{\infty}{[K(\alpha'/\tilde{b})^2 a^4 u^4-1]\over (1+Ka^4 u^4)}
            (d\tilde{x}_{1}\wedge d\tilde{x}_{3}), \nonumber \\
A_{[2]} &=& {1\over g_{s}}\left({\tilde{b}\over \alpha'}\right)(\alpha'^2 R^4 u^4-1)
            (d\tilde{t}\wedge d\tilde{x}_{2}), \\
A_{[4]} &=& {\hat{h}\over g_{s}}(\alpha'^2 R^4 u^4-1)
            (d\tilde{t}\wedge d\tilde{x}_{1}\wedge d\tilde{x}_{2}\wedge d\tilde{x}_{3}), 
            \nonumber \\
e^{2\phi} &=& \hat{g}^2_{s}\hat{h} \nonumber
\eeqa
where $K$ and $B_{\infty}$ are as given above but now $\hat{h}=1/(1+Ka^4 u^4)$. \\
The decoupling limit of these extremal $(D1-D3)$ bound state solutions with a
superimposed pp-wave given above is the main result we would like to report in this work.
Namely, in the spirit of $K_{(p+2)}/CFT_{(p+1)}$ correspondence that we discussed earlier
in the introduction, we propose that the decoupling limit of the extremal solutions given
above constitute the dual supergravity description of the SYM theory in the 
infinitely-boosted frame on noncommutative 4-dimensional manifold. And in the same spirit,
we expect that the decoupling limit of the non-extremal solutions should be the dual
supergravity description of the noncommutative SYM theory in the infinitely-boosted frame
at {\it finite temperature}. At this point, it seems relevant to point out the interesting role
played by the presence of the pp-wave parallely-intersecting with the $(D1-D3)$ bound state.
To do so, recall first that the solution construction corresponding to option (A) involves,
when obtaining the $(D1-D3)$ bound state from the $D2$-brane solution via the so-called
BMM $T$-duality prescription, the rotation in a plane containing a {\it spectator} direction
(with respect to the superimposed pp-wave), while the one corresponding to option (B) involves
the rotation in the other plane containing the {\it boost} (i.e., wave propagation) direction.
These rather technically-looking choices corresponding to the two inequivalent options in the
solution construction procedure, however, turn out to lead to physically interesting
consequences. Namely, the decoupling limit of the solution corresponding to option (A) is expected
to be the dual gravity description of the SYM theory in the infinitely-boosted frame on a
manifold with one noncommutative hypersurface, $(\tilde{x}_{2}-\tilde{x}_{3})$ plane.
In contrast, that corresponding to option (B) is supposed to be the dual gravity description
of the same gauge theory, this time on a manifold with two noncommutative hypersurfaces,
$(\tilde{t}-\tilde{x}_{3})$, $(\tilde{x}_{1}-\tilde{x}_{3})$ planes. In other words, one ends
up with both ``time-space'' and ``space-space'' noncommutativity in option (B) in contrast to
option (A) where only ``space-space'' noncommutativity is present. The essential reason that
underlies this emergence of ``time-space'' noncommutativity already has been discussed earlier.
Here, we stress that on purely technical side, intersecting the $(D1-D3)$ bound state parallely
with a gravitational pp-wave turns out to provide yet another way of generating the
``time-space'' noncommutativity in its dual SYM theory different from those suggested in the
literature in the absence of the pp-wave. \\
Now, other comments concerning the decoupling limit of these extremal $(D1-D3)$ bound state
with a superimposed pp-wave are in order : \\

(i) These extremal solutions all approach $K_{5}\times S^{5}$ (with $K_{5}$ denoting the
5-dimensional generalisation of the ``Kaigorodov'' metric) for small $u$, which corresponds to
the IR regime of the dual SYM theory in the infinitely-boosted frame since ``$u$'' plays the role
of energy scale on the gauge theory side. This is indeed what one would naturally expect since
the noncommutative SYM theory should reduce to the ordinary (commutative) SYM theory at long
distances. And the solutions start to deviate from the $K_{5}\times S^{5}$ solution roughly
at $u\sim 1/a$ (with $a = R\sqrt{\tilde{b}}$ carrying the dimension of the length), namely
at a distance scale of order $a = R\sqrt{\tilde{b}}$. As Maldacena and Russo \cite{mr} pointed out,
for large `tHooft coupling, i.e., for $\lambda = g^2_{YM}N = 4\pi g_{s}N = R^4\sim
{\rm large}$, this clearly is greater than the naively expected distance scale of 
$L\sim \sqrt{\tilde{b}}$. \\

(ii) We now turn to the behavior of these solutions on the other asymptotic boundary at
$u\to \infty$. Unlike the case in which the $NSNS$ $B$-field is absent, the solutions exhibit
some peculiar features. For instance, as this boundary is approached, the physical (proper)
size of the noncommutative directions (i.e., $\tilde{x}_{2}-\tilde{x}_{3}$ directions) shrink
(in string frames), since $\hat{h} = 1/(1+a^4 u^4) \sim 1/u^4$ as $u\to \infty$ for the solution
corresponding to option (A). Interestingly, however, this is not the case for the solution
corresponding to option (B) since there 
$\hat{h} = 1/(1+Ka^4 u^4) = [1+Q_{2}(a/\alpha' R^2)^4]^{-1}$ and hence it is independent of $u$
in the decoupling limit. Namely for the solution corresponding to option (B), the physical size
of both the commuting and noncommuting directions exhibits essentially the same (growing)
behavior as the $u\to \infty$ boundary is approached. And for the solution corresponding to
option (A), this shrinking behavior of the physical size of the noncommuting directions may lead
to the danger of encountering the curvature singularity (in string metric) as $u\to \infty$
since the type of scaling isometry near this boundary that exists in the absence of the pp-wave
content, 
\beq
\tilde{x}_{0,1}\to \lambda^{-1}\tilde{x}_{0,1}, ~~~\tilde{x}_{2,3}\to \lambda \tilde{x}_{2,3},
~~~u\to \lambda u \nonumber
\labell{D5actexp}
\eeq
noticed by Maldacena and Russo \cite{mr} simply does not exist for the case at hand when the 
pp-wave content is present. \\

(iii) We now briefly comment on the nature of supersymmetry and some duality owned by these
solutions. The transverse 5-sphere is still round and hence possesses $SO(6)$-isometry which
corresponds on the dual gauge theory side to the $SU(4)$ $R$-symmetry of the ${\cal N}=4$ SUSY 
algebra. And the fact that this $SO(6)$-isometry is not contaminated even under the introduction 
of the noncommutativity implies that the SUSY is not further broken by the noncommutativity either.
Next, the electric 5-form $RR$ field strength of the electric solution is apparently not the
same in form as the magnetic 5-form $RR$ field strength of the magnetic solution, i.e.,
$F^{e}_{[5]} \neq F^{m}_{[5]} = ^{*}F^{e}_{[5]}$ in eqs.(27) and (32). This is due to
the presence of the $NSNS$ $B$-field (leading to the $(D1-D3)$ bound state) and the gravitational
pp-wave propagating on the brane and implies that this particular type IIB supergravity 
solution is not self-dual under $S$-duality. \\

\section{Summary and Discussion}

To summarize, in the present work, we attempted to explore the mechanism of non-locality in the
noncommutative SYM theory in the infinitely-boosted frame especially at strong coupling from
the dual supergravity description in terms of the extremal $(D1-D3)$ bound state solution with
a superimposed pp-wave. One may naturally expect that when the effect of non-locality of order,
say, $a=R\sqrt{\tilde{b}}=(g^2_{YM}N\tilde{b}^2)^{1/4}$ is turned on, the dynamics at length
scales larger than $a$ would be unaffected while that at length scales smaller than $a$ would
be drastically changed. From the gravity decoupling limit of the dual supergravity solution, 
we have actually confirmed this intuitive expectation. That is, for small $u$, the solutions
all reduce to $K_{5}\times S^5$ geometry confirming our expectation that the IR dynamics of the
dual gauge theory should be unaffected by the noncommutativity while as $u\to \infty$, the
solutions start to deviate significantly from $K_{5}\times S^5$ limit indicating that the
UV dynamics of the dual gauge theory would be heavily distorted by the effect of noncommutativity.\\

Nevertheless, aside from our attempt to study it using the $K_{(p+2)}/CFT_{(p+1)}$ correspondence,
the noncommutative SYM theory in the infinitely-boosted frame itself does not seem to have
been studied in great detail. Thus it might be challenging to work in this direction as well. 
The commutative boundary CFT in the infinitely-boosted frame, on the other hand, has been examined
by Brecher, Chamblin and Reall \cite{bcr} in some detail. Thus it seems worth summarizing the 
results of their study here. They also started by noting that in the spirit of gauge/gravity
correspondence, it is natural to conjecture that string theory in the Kaigorodov spacetime is
dual to some CFT in the infinitely-boosted frame. Since the momentum density was held fixed in
the gravity decoupling limit, however, there is a non-zero background momentum density present.
And this background momentum density, in turn, breaks the conformal symmetry group of the 
boundary field theory down to some smaller group. They showed that actually the isometries of
the Kaigorodov spacetime have a natural interpretation as this subgroup of the conformal group
that leaves the background momentum density invariant. They then attempted the computation of
2-point functions of field operators in the boundary theory. As is well-known, when conformal
symmetry is exact, the 2-point functions of CFT operators are completely determined. For the
case under consideration when the conformal invariance is partly broken, the dilatation
symmetry still persists and it allows to constrain the form of 2-point functions. As a result,
they demonstrated that this surviving symmetry determines the scalar 2-point function up to an
arbitrary function of one variable. Moreover, this 2-point function turned out to be independent
of the background momentum density and this point has been attributed to a large $N$ effect. 
In association with the context of the present work in which the gravity duals to the 
noncommutative boundary CFT in the infinitely-boosted frame has been developed, then, one
might wish to add the noncommutativity content to the type of analysis performed in \cite{bcr}
to eventually study the corresponding dual CFT. Technically, it can be achieved by properly
combining the works \cite{mr} and \cite{bcr}. And this will be left for a serious future study.

\section*{Acknowledgments}

This work was financially supported by the BK21 Project of the Korean Government.

\appendix
\section{$M2$-brane in $D=11$ SUGRA with a superimposed pp-wave}

Consider the (bosonic sector of) $D=11$ SUGRA with action
\begin{eqnarray}
S_{11} &=& \int d^{11}x\sqrt{g}\left[R - {1\over 2\times 4!}F^{2}_{[4]}\right]
        - {1\over 6}\int A_{[3]}\wedge F_{[4]}\wedge F_{[4]}, \nonumber \\
A_{[3]} &=& {1\over 3!}A_{MNP}dx^{M}\wedge dx^{N}\wedge dx^{P}, ~~~F_{[4]}=dA_{[3]} 
\end{eqnarray}
and the associated classical field equations
\begin{eqnarray}
&&R_{MN} = {1\over 2\times 3!}\left[F_{MPQR}F_{N}^{PQR} - {1\over 12}g_{MN}F^{2}\right], \nonumber \\
&&{1\over \sqrt{g}}\partial_{M}\left[\sqrt{g}F^{MPQR}\right] = 0.
\end{eqnarray}
Then the non-extremal $M2$-brane solution to this classical field equations is given by
\beqa
ds^2_{11} &=& H^{-2/3}\left[-fdt^2 + dx^2_{1} 
              + dx^2_{2}\right] + H^{1/3}\left[f^{-1}dr^2 + r^2 d\Omega^2_{7}\right], \nonumber \\
A_{[3]} &=& \coth \mu_{1}(H^{-1}-1) (dt\wedge dx_{1}\wedge dx_{2}) ~~~{\rm with} \\
H(r) &=& 1+{Q_{1}\over r^6}, ~~~f(r) = 1-{\mu\over r^6}, 
~~~Q_{1} = \mu \sinh^2\mu_{1} = (2^{5} \pi^{2} N)l^{6}_{p}, ~~~\mu = k\kappa^{4/3}_{11} = 2m \nonumber
\eeqa
where $Q_{1}$ is the usual (electric) $RR$ charge and $\kappa_{11}$ and $l_{p}$ denotes the 
11-dimensional gravitational constant and Planck length respectively and $N$ is the number of
coincident branes and lastly $m$ denotes the (ADM) mass density of the brane.  
And the horizon of this black $M2$-brane is located at 
$r_{+}=\mu^{1/6}=k^{1/6}\kappa^{2/9}_{11}$ along the transverse radial direction. \\
Now, note that in the presence, say, of an electric charge (and its field), one way of
generating the electromagnetic wave is to go to an {\it infinitely-boosted} Lorentz frame,
i.e., a Lorentz frame moving at the speed of light with respect to the charge. 
In a similar manner, in the presence of (some form of) a mass, like $M2$-brane itself given
above, one way of generating the gravitational pp-wave would be to make a transit to an
infinitely-boosted Lorentz frame. Thus in order to construct a non-extremal $M2$-brane
solution with a superimposed gravitational pp-wave, we consider performing a Lorentz boost 
on the non-extremal $M2$-brane solution given above in the $(t,~x_{1})$ plane
\beqa
t' &=& (\cosh \mu_{2})t + (\sinh \mu_{2})x_{1}, \\
x'_{1} &=& (\sinh \mu_{2})t + (\cosh \mu_{2})x_{1} \nonumber
\eeqa
where $\gamma = (1-\beta^2)^{-1/2} = \cosh \mu_{2}$, $\beta \gamma = \sinh \mu_{2}$,
and thus $\beta = \sinh \mu_{2}/\gamma = \tanh \mu_{2}$. Note, as mentioned earlier in the
introduction, that this Lorentz boost becomes singular, i.e., becomes an infinite boost in
the extremal limit where $\mu_{2}\to \infty$. Upon this Lorentz boost, then,
the part of the $M2$-brane worldvolume metric becomes
\beqa
&&[-f(r)dt^2 + dx^2_{1} + dx^2_{2}]  \\
&&= -f(r)(\cosh \mu_{2}dt'-\sinh \mu_{2}dx'_{1})^2 
  + (\cosh \mu_{2}dx'_{1}-\sinh \mu_{2}dt')^2 + dx^2_{2} \nonumber \\
&&= -K^{-1}f(r)dt'^2 + K[dx'_{1} + \coth \mu_{2}(K^{-1}-1)dt']^2 + dx^2_{2} \nonumber \\
&&{\rm with} ~~~K(r) = 1 + {Q_{2}\over r^6}, ~~~Q_{2} = \mu \sinh^2\mu_{2} \nonumber
\eeqa
and where $Q_{2}$ is a new parameter representing the momentum along $x'_{1}$-direction.
Next, it is straightforward to see that under this Lorentz boost, the $RR$ tensor field 
(and hence its field strength) remains the same, namely
\beqa
A'_{t'x'_{1}x_{2}} &=& \left({\partial t\over \partial t'}{\partial x_{1}\over \partial x'_{1}}
- {\partial x_{1}\over \partial t'}{\partial t\over \partial x'_{1}}\right) A_{tx_{1}x_{2}} \nonumber \\
&=& (\cosh^2\mu_{2} - \sinh^2\mu_{2})A_{tx_{1}x_{2}} = A_{tx_{1}x_{2}} ~~~{\rm and}   \\
dt'\wedge dx'_{1}\wedge dx_{2} &=& (\cosh^2\mu_{2} - \sinh^2\mu_{2}) dt\wedge dx_{1}\wedge dx_{2}
= dt\wedge dx_{1}\wedge dx_{2} ~~~{\rm thus} \nonumber \\
A_{[3]} &=& A_{tx_{1}x_{2}} dt\wedge dx_{1}\wedge dx_{2}
= A'_{t'x'_{1}x_{2}} dt'\wedge dx'_{1}\wedge dx_{2}. \nonumber
\eeqa
Thus putting these results altogether, one can conclude that upon the Lorentz boost 
in the $(t,~x_{1})$ plane, the non-extremal $M2$-brane solution goes over to the non-extremal
$M2$-brane solution with a superimposed gravitational pp-wave given by (henceforth, we shall
drop the primes on $t$ and $x_{1}$ coordinates)
\beqa
ds^2_{11} &=& H^{-2/3}\left[-K^{-1}fdt^2 + K \{dx_{1}+\coth \mu_{2}(K^{-1}-1)dt\}^2 
              + dx^2_{2}\right] + H^{1/3}\left[f^{-1}dr^2 + r^2 d\Omega^2_{7}\right], \nonumber \\
A_{[3]} &=& \coth \mu_{1}(H^{-1}-1) (dt\wedge dx_{1}\wedge dx_{2}) ~~~{\rm with} \\
H(r) &=& 1+{Q_{1}\over r^6}, ~~~K(r) = 1+{Q_{2}\over r^6},  ~~~f(r) = 1-{\mu\over r^6}, \nonumber \\
Q_{1} &=& \mu \sinh^2\mu_{1}, ~~~Q_{2} = \mu \sinh^2\mu_{2}, ~~~\mu = k\kappa^{4/3}_{11}. \nonumber
\eeqa
Next, the extremal $M2$-brane with a superimposed gravitational pp-wave amounts to the limiting
case when
\beq
\mu\to 0, ~~~\mu_{1}, ~\mu_{2}\to \infty ~~~{\rm with} ~~~Q_{1} = \mu \sinh^2\mu_{1},
~~~Q_{2} = \mu \sinh^2\mu_{2} ~~~{\rm kept ~~fixed}
\eeq
then the solution above becomes
\beqa
ds^2_{11} &=& H^{-2/3}\left[-K^{-1}dt^2 + K \{dx_{1}+(K^{-1}-1)dt\}^2 + dx^2_{2}\right]
              + H^{1/3}\left[dr^2 + r^2 d\Omega^2_{7}\right], \nonumber \\
A_{[3]} &=& (H^{-1}-1) (dt\wedge dx_{1}\wedge dx_{2}). 
\eeqa
Finally, upon introducing the retarded $(u)$ and the advanced $(v)$ null coordinates
\beq
u = x_{1} - t,  ~~~v = x_{1} + t \nonumber
\eeq
the extremal solution takes the form
\beq
ds^2_{11} = H^{-2/3}(r)[dudv + (K-1)du^2 + dx^2_{2}] + H^{1/3}(r)[dr^2 + r^2d\Omega^2_{7}].
\eeq
Obviously, in this extreme limit which corresponds to the infinite Lorentz boost case, the metric
on the $M2$-brane worldvolume for $r = const.$ ;
\beq
[dudv + (K-1)du^2 + dx^2_{2}] ~~~{\rm with} ~~~(K-1) = {Q_{2}\over r^6} \nonumber
\eeq
does indeed represent a gravitational wave propagating in $x_{1}$-direction.

\section{What is the gravitational ``pp-wave''?}

By definition, a vacuum spacetime is a {\it plane-fronted gravitational waves} provided it
contains a ``shear-free'' congruence of null geodesics (with tangent $k^{\alpha}$) and
provided it admits ``plane wave surfaces'' (i.e., spacelike 2-surfaces orthogonal to 
$k^{\alpha}$). And because of the existence of plane wave surfaces, the expansion and twist
(rotation) must vanish as well. The best-known subclass of these waves are {\it plane-fronted
gravitational waves with parallel rays} (``pp-waves'') which are defined by the condition that
the null vector $k^{\alpha}$ is covariantly constant, $\nabla_{\beta}k_{\alpha}=0$.  \\
Generally, for the null vector $k^{\alpha}$ tangent to null geodesic congruence, 
$\nabla_{\beta}k_{\alpha}$ can be decomposed as
\beq
\nabla_{\beta}k_{\alpha} = \sigma_{\alpha\beta} + \omega_{\alpha\beta} + {1\over 3}\theta
h_{\alpha\beta} - a_{\alpha}k_{\beta}
\eeq
where $h_{\alpha\beta} = g_{\alpha\beta} + k_{\alpha}k_{\beta}$ is the metric induced on the
hypersurfaces $\Sigma$ orthogonal to $k^{\alpha}$ and
\beqa
\theta &\equiv& h^{\alpha\beta}\nabla_{\beta}k_{\alpha} = \nabla^{\alpha}k_{\alpha}, \nonumber \\
\sigma_{\alpha\beta} &\equiv& {1\over 2}(h^{\mu}_{\beta}\nabla_{\mu}k_{\alpha} +
h^{\mu}_{\alpha}\nabla_{\mu}k_{\beta}) - {1\over 3}\theta h_{\alpha\beta} =
\nabla_{(\alpha}k_{\beta)} - {1\over 3}\theta h_{\alpha\beta}, \nonumber \\
\omega_{\alpha\beta} &\equiv& {1\over 2}(h^{\mu}_{\beta}\nabla_{\mu}k_{\alpha} -
h^{\mu}_{\alpha}\nabla_{\mu}k_{\beta}) = \nabla_{[\alpha}k_{\beta]}, \\
a_{\alpha} &\equiv& k^{\beta}\nabla_{\beta}k_{\alpha} \nonumber
\eeqa
are the expansion, shear, twist and acceleration, respectively, of the null
geodesic congruence. Thus if the expansion, shear, twist and acceleration all vanish, then
$\nabla_{\beta}k_{\alpha}=0$. In suitable null coordinates $(u,~v,~\xi,~\bar{\xi})$ such that
\beq
k_{\alpha} = \partial_{\alpha}u, ~~~k^{\alpha} = \left(\partial/\partial v\right)^{\alpha}
\eeq
the metric representing the gravitational pp-wave is given by
\beqa
ds^2 &=& dudv + H(u,~\xi,~\bar{\xi})du^2 + (dx^2_{2} + dx^2_{3}) \nonumber \\
     &=& dudv + H(u,~\xi,~\bar{\xi})du^2 + d\xi d\bar{\xi} 
\eeqa
where $H$ is a real function of $u$ and $\xi$ which spans the wave 2-surfaces 
$u = const.$, $v = const.$ The vacuum Einstein field equations imply
$2H = f(u,~\xi) + \bar{f}(u,~\bar{\xi})$ with $f$ being an arbitrary function of $u$ analytic
in $\xi$. In general, the gravitational pp-waves have only the single isometry generated by the
Killing vector $k^{\alpha} = \left(\partial/\partial v\right)^{\alpha}$.

\end{document}